\documentstyle[aps]{revtex}

\begin{document}
\title{The classical regime of a quantum universe obtained through a functional
method.}
\author{Mario Castagnino}
\address{Instituto de Astronom\'{\i}a y F\'{\i}sica del Espacio\\
Casilla de Correos 67, Sucursal 28.\\
1482 Buenos Aires, Argentina.}
\maketitle

\begin{abstract}
The functional method, introduced to deal with systems endowed with a
continuous spectrum, is used to study the problem of decoherence and
correlations in a simple cosmological model.
\end{abstract}

\section{Introduction.}

One of the most important problems of theoretical physics in the last years
was to answer the question: How and in what circumstances a quantum system
becomes classical ? \cite{WyZ}. In spite of the great effort made by the
physicists to find the answer, the problem is still alive \cite{Giuliani}
and we are far from a complete understanding of many of its most fundamental
features. In fact the most developed and sophisticated theory on the
subject: histories decoherence is not free of strong criticisms \cite{Dowker}%
.

Nevertheless there is an almost unanimous opinion that the classical regime
is produced by two phenomena:

i. {\it Decoherence,} that in quantum systems, restore the boolean statistic
typical of quantum mechanics and

ii.-{\it Correlations, }that circumvent the uncertainty relation at the
macroscopic level.

But the techniques to deal with these two phenomena are not yet completely
developed. One of the main problems is to find a proper and unambiguous
definition of the, so called, {\it pointer basis,} where, decoherence takes
place.

Our contribution to solve this problem is based in old ideas of Segal \cite
{Segal} and van Howe \cite{van Howe}, reformulated by Antoniou et al. \cite
{Antoniou}. We have developed these ideas in papers \cite{LauraA} and \cite
{LauraE} where we have shown how the Riemann-Lebesgue theorem can be used to
prove the destructive interference of the off-diagonal terms of the state
density matrix yielding decoherence. Using this technique we have found
decoherence and correlations in simple quantum systems \cite{LauraF} where
we have defined the pointer basis in an unambiguous way. \footnote{%
The relation of our method with the histories decoherence is studied in
paper \cite{LauraF}. They turn out to be equivalent, but in our method the
pointer basis is better defined.}.

On the other hand, the appearance of a classical universe in quantum gravity
models is the cosmological version of the problem we are discussing. Then,
decoherence and correlations must also appear in the universe \cite{Cast
decoher}. In this paper, using our method, we will solve this problem in a
simple quantum-cosmological model and we will find:

i.-Decoherence in all the dynamical variables and in a well defined pointer
basis.

ii.-Correlations, in such a way that the Wigner function $F_{W*}$ of the
asymptotic diagonal matrix $\rho _{*}$ can be expanded as: 
\begin{equation}
F_{W*}(x,p)=\int p_{\{l\}[{\bf a}]}F_{W\{l\}[{\bf a}]}(x,p)d\{l\}d[{\bf a}]
\label{1.1}
\end{equation}
where $F_{W\{l\}[{\bf a}]}$ is a classical density strongly peaked \footnote{%
Precisely: peaked as allowed by the uncertainty principle.} in a trajectory
defined by the initial conditions {\bf a} and the momenta $l$ and $p_{\{l\}[%
{\bf a}]}$ the probability of each trajectory. As the limit of quantum
mechanics is not classical mechanics but classical statistical mechanics
this is our final result: The density matrix is translated in a classical
density, via a Wigner function, and it is decomposed as a sum of densities
peaked around all possible classical trajectories, each one of these
densities weighted by their own probability.

Thus our quantum density matrix behaves in its classical limit as a
statistical distribution among a set of classical trajectories. Similar
results are obtained in papers \cite{Zouppas} and \cite{Polarsky}.

\section{The model.}

Let us consider the flat Roberson-Walker universe (\cite{Juanpa}, \cite{Cast}%
) with a metric: 
\begin{equation}
ds^2=a^2(\eta )(d\eta ^2-dx^2-dy^2-dz^2)  \label{2.1}
\end{equation}
where $\eta $ is the conformal time and $a$ the scale of the universe. Let
us consider a free neutral scalar field and let us couple this field with
the metric, with a conformal coupling ($\xi =\frac 16)$. The total action
reads $S=S_g+S_f$ $+S_i$ and the gravitational action is: 
\begin{equation}
S_g=M^2\int d\eta [-\frac 12\stackrel{\bullet }{a}^2-V(a)]  \label{2.2}
\end{equation}
where $M$ is the Planck mass, $\stackrel{\bullet }{a}=da/d\eta ,$ and the
potential $V$ contains the a cosmological constant term and eventually the
contribution of some form of classical mater. We suppose that $V$ has a
bounded support $0\leq a\leq a_1.$ We expand the field $\Phi $ as: 
\begin{equation}
\Phi (\eta ,{\bf x})=\int f_{{\bf k}}e^{-i{\bf k\cdot x}}d{\bf k}
\label{2.3}
\end{equation}
where the components of ${\bf k}$ are three continuous variables.

The Wheeler De-Witt equation for this model reads: 
\begin{equation}
H\Psi (a,\Phi )=(h_g+h_f+h_i)\Psi (a,\Phi )=0  \label{2.4}
\end{equation}
where: 
\[
h_g=\frac 1{2M^2}\partial _a^2+M^2V(a) 
\]
\[
h_f=-\frac 12\int (\partial _{{\bf k}}^2-k^2f_{{\bf k}}^2)d{\bf k} 
\]

\begin{equation}
h_i=\frac 12m^2a^2\int f_{{\bf k}}^2d{\bf k}  \label{2.5}
\end{equation}
being $m$ is the mass of the scalar field, ${\bf k}/a$ is the linear
momentum of the field, and $\partial _{{\bf k}}$ =$\partial /\partial f_{%
{\bf k}}.$

We can now go to the semiclassical regime using the WKB method (\cite{Hartle}%
), writing $\Psi (a,\Phi )$ as: 
\begin{equation}
\Psi (a,\Phi )=\exp [iM^2S(a)]\chi (a,\Phi )  \label{2.6}
\end{equation}
and expanding $S$ and $\chi $ as: 
\begin{equation}
S=S_0+M^{-1}S_1+...,\qquad \chi =\chi _0+M^{-1}\chi _1+...  \label{2.7}
\end{equation}
To satisfy eq. (\ref{2.4}) at the order $M^2$ the principal Jacobi function $%
S(a)$ must satisfy the Hamilton-Jacobi equation: 
\begin{equation}
\left( \frac{dS}{da}\right) ^2=2V(a)  \label{2.8}
\end{equation}
We can now define the (semi)classical time as a parameter $\eta =\eta (a)$
such that: 
\begin{equation}
\frac d{d\eta }=\frac{dS}{da}\frac d{da}=\pm \sqrt{2V(a)}\frac d{da}
\label{2.9}
\end{equation}
The solution of this equation is $a=\pm F(\eta ,C),$ where $C$ is an
arbitrary integration constant. Different values of this constant and of the 
$\pm $ sign give different classical solutions for the geometry.

Then, in the next order of the WKB expansion, the Schroedinger equation
reads: 
\begin{equation}
i\frac{d\chi }{d\eta }=h(\eta )\chi  \label{2.10}
\end{equation}
where: 
\begin{equation}
h(\eta )=h_f+h_i(a)  \label{2.11}
\end{equation}
precisely: 
\begin{equation}
h(\eta )=-\frac 12\int \left[ -\frac{\partial ^2}{\partial f_{{\bf k}}^2}%
+\Omega _{{\bf k}}^2(a)f_k^2\right] d{\bf k}  \label{2.12}
\end{equation}
where: 
\begin{equation}
\Omega _\varpi ^2=m^2a^2+k^2=m^2a^2+\varpi  \label{2.13}
\end{equation}
where $\varpi =k^2$ and $k=|{\bf k|.}$ So the time dependence of the
hamiltonian comes from the function $a=a(\eta ).$

Let us now consider a scale of the universe such that $a_{out}\gg a_1$. In
this region the geometry is almost constants. Therefore we have an adiabatic
final vacuum $|0\rangle $ and adiabatic creation and annihilation operators $%
a_{{\bf k}}^{\dagger }$ and $a_{{\bf k}}.$ Then $h=h(a_{out})$ reads: 
\begin{equation}
h=\int \Omega _\varpi a_{{\bf k}}^{\dagger }a_{{\bf k}}d{\bf k}  \label{2.14}
\end{equation}

We can now consider the Fock space and a basis of vectors: 
\begin{equation}
|{\bf k}_1,{\bf k}_2,...,{\bf k}_n,...\rangle \cong |\{k{\bf \}\rangle =}a_{%
{\bf k}_1}^{\dagger }a_{{\bf k}_2}^{\dagger }...a_{{\bf k}_n}^{\dagger
}...|0\rangle  \label{2.15}
\end{equation}
where we have called $\{k{\bf \}}$ to the set ${\bf k}_1,{\bf k}_2,...,{\bf k%
}_n,...$ The vectors of this basis are eigenvectors of $h:$%
\begin{equation}
h|\{k{\bf \}\rangle =}\omega |\{k{\bf \}\rangle }  \label{2.16}
\end{equation}
where: 
\begin{equation}
\omega =\sum_{{\bf k\in \{k\}}}\Omega _\varpi =\sum_{{\bf k\in \{k\}}%
}(m^2a_{out}^2+\varpi )^{\frac 12}  \label{2.17}
\end{equation}
We can now use this energy to label the eigenvector as: 
\begin{equation}
|\{k{\bf \}\rangle =}|\omega ,[{\bf k]\rangle }  \label{2.18}
\end{equation}
where $[{\bf k]}$ is the remaining set of labels necessary to define the
vector unambiguously. $\{|\omega ,[{\bf k]\rangle \}}$ is obviously an
orthonormal basis so eq. (\ref{2.14}) reads: 
\begin{equation}
h=\int \omega |\omega ,[{\bf k]\rangle }\langle \omega ,[{\bf k]|}d\omega d[%
{\bf k]}  \label{2.19}
\end{equation}
In the next section we will write this equation using a shorthand notation
as: 
\begin{equation}
h=\int \omega |\omega {\bf \rangle }\langle \omega {\bf |}d\omega
\label{2.20}
\end{equation}
The dynamical variables $[{\bf k]}$ will reappear in section IV.

\section{Energy decoherence.}

As we are dealing with a system with a continuous spectrum in $\omega ,$
some care must be taken. If not the mathematical manipulations can contain
multiplication of distribution and yield infinite meaningless results. In
order to deal with this problem and to always work with usual functions (not
distribution) a {\it functional method }was introduced in papers \cite
{LauraA} and \cite{LauraE}, that we will now review. The method was used to
show the decoherence and the existence of correlations in ordinary quantum
mechanical systems \cite{LauraF} and we will use it in our problem.

The physical basis of the method is the following: The states of the
universe are only known and measure trough a measurement process where a
space of observables ${\cal O}$ is used. For any observable $O\in {\cal O}$
we can only measure the mean value of this $O$ in a state $\rho ,$ namely:

\begin{equation}
\langle O\rangle _\rho =Tr(\rho ^{\dagger }O)  \label{3.1}
\end{equation}
Then we can consider that the states are linear functionals over the space
of observables and write: 
\begin{equation}
\langle O\rangle _\rho =\rho [O]=(\rho |O)  \label{3.2}
\end{equation}
Of course, the states would be endowed with extra some properties, so we
will define a convex set of observables ${\cal S\subset O}^{\prime }$ ,
being this last space the dual of ${\cal O}$ .

It is logical to ask that the hamiltonian $h$ would be contained in the
space of observables ${\cal O}$ , then the observables must be defined
generalizing eq. (\ref{2.20}). This generalization, already used in papers 
\cite{LauraA} and \cite{LauraE}, reads:

\[
O=\int O_\omega |\omega \rangle \langle \omega |d\omega +\int \int O_{\omega
\omega ^{\prime }}|\omega \rangle \langle \omega ^{\prime }|d\omega d\omega
^{\prime }= 
\]
\begin{equation}
\int O_\omega |\omega )d\omega +\int \int O_{\omega \omega ^{\prime
}}|\omega ,\omega ^{\prime })d\omega d\omega ^{\prime }  \label{3.3}
\end{equation}
where we have introduced a $basis$ $\{|\omega ),|\omega ,\omega ^{\prime
})\} $of space ${\cal O}$ defined as: 
\begin{equation}
|\omega )=|\omega \rangle \langle \omega |,\qquad |\omega ,\omega ^{\prime
})=|\omega \rangle \langle \omega ^{\prime }|  \label{3.4}
\end{equation}
The terms $O_\omega $ can be considered as the (singular) diagonal terms,
while the terms $O_{\omega \omega ^{\prime }\text{ }}$ can be considered as
the (regular) off-diagonal terms.

We can now define the $cobasis$ $\{(\omega |,(\omega ,\omega ^{\prime }|\}$
of space ${\cal O}$ (namely the basis of space ${\cal S\subset O}^{\prime }$%
), that obviously satisfies the equations:$\;$%
\begin{equation}
(\omega |\omega ^{\prime })=\delta (\omega -\omega ^{\prime }),\qquad
(\omega ,\omega ^{\prime \prime }|\omega ^{\prime },\omega ^{\prime \prime
\prime })=\delta (\omega -\omega ^{\prime })\delta (\omega ^{\prime \prime
}-\omega ^{\prime \prime \prime })  \label{3.5}
\end{equation}
and all other $(.|.)=0.$

Then if $\rho \in {\cal S}$ it can be expanded as: 
\begin{equation}
\rho =\int \rho _\omega (\omega |d\omega +\int \int \rho _{\omega \omega
^{\prime }}(\omega ,\omega ^{\prime }|d\omega d\omega ^{\prime }  \label{3.6}
\end{equation}
where $\rho _\omega \geq 0,$ $\rho _{\omega \omega ^{\prime }}=\rho _{\omega
^{\prime }\omega }^{*}.$ Moreover, the ordinary functions $O_\omega
,O_{\omega \omega ^{\prime },}\rho _\omega ,$ and $\rho _{\omega \omega
^{\prime }},$ must be endowed with certain properties in order to make all
the equations of the formalism well defined. These properties are listed in 
\cite{LauraA} and they are assumed in this paper. Then: 
\begin{equation}
(\rho |O)=\int \rho _\omega O_\omega d\omega +\int \int \rho _{\omega \omega
^{\prime }}O_{\omega ^{\prime }\omega }d\omega d\omega ^{\prime }
\label{3.7}
\end{equation}
and from eq. (\ref{2.10}): 
\begin{equation}
(\rho (\eta )|O)=\int \rho _\omega O_\omega d\omega +\int \int \rho _{\omega
\omega ^{\prime }}O_{\omega ^{\prime }\omega }e^{i(\omega -\omega ^{\prime
})\eta }d\omega d\omega ^{\prime }  \label{3.8}
\end{equation}
Then, when $\eta \rightarrow \infty ,$ essentially from the Riemann-Lebesgue
theorem (see \cite{LauraA} for details), we have: 
\begin{equation}
\lim_{\eta \rightarrow \infty }(\rho (\eta )|O)=\int \rho _\omega O_\omega
d\omega =(\rho _{*}|O)  \label{3.9}
\end{equation}
where: 
\begin{equation}
(\rho _{*}|=\int \rho _\omega (\omega |d\omega  \label{3.10}
\end{equation}
is the equilibrium time-asymptotic state, which only contains the diagonal
term. So we have proved the existence of decoherence in the energy.

\section{Decoherence in the other dynamical variables.}

If we reintroduce the other dynamical variables in eq. (\ref{3.10}) we
obtain: 
\begin{equation}
(\rho _{*}|=\int \rho _{\omega [{\bf k][k}^{\prime }]}(\omega ,[{\bf k],[k}%
^{\prime }]|d\omega d[{\bf k]d[k}^{\prime }]  \label{4.1}
\end{equation}
where \{$(\omega ,[{\bf k],[k}^{\prime }]|,(\omega ,\omega ^{\prime },[{\bf %
k],[k}^{\prime }]\}$ is the cobasis \{$(\omega |,(\omega ,\omega ^{\prime
}|\}$ but now showing the hidden $[{\bf k].}$

Let us observe that if we would use polar coordinates for ${\bf k}$ eq.(\ref
{2.3}) reads:

\begin{equation}
\Phi (x,n)=\int \sum_{lm}\phi _{klm}dk  \label{4.2}
\end{equation}
where:

\begin{equation}
\phi _{klm}=f_{k,l}(\eta ,r)Y_m^l(\theta ,\varphi )  \label{4.3}
\end{equation}
where $k$ is a continuous variable, $l=0,1,...,;$ $m=-l,...,l;$ and $Y$ are
spherical harmonic functions. So the indices $k,l,m$ contained in the symbol 
${\bf k}$ are partially discrete and partially continuous.

As $\rho _{*}^{\dagger }=\rho _{*}$ then $\rho _{\omega [{\bf k}^{\prime }%
{\bf ][k}]}^{*}=$ $\rho _{\omega [{\bf k][k}^{\prime }]}$ and therefore a
set of vectors $\{|\omega ,[{\bf l}]\rangle \}$ exists such that: 
\begin{equation}
\int \rho _{\omega [{\bf k][k}^{\prime }]}|\omega ,[{\bf l}]\rangle _{[{\bf k%
}^{\prime }]}d[{\bf k}^{\prime }]=\rho _{\omega [{\bf l]}}|\omega ,[{\bf l}%
]\rangle _{[{\bf k]}}  \label{4.4}
\end{equation}
namely \{$|\omega ,[{\bf l}]\rangle \}$ is the eigenbasis of the operator $%
\rho _{\omega [{\bf k][k}^{\prime }]}.$ Then $\rho _{\omega [{\bf l]}} $ can
be considered as an ordinary diagonal matrix in the discrete indices like
the $l$ and the $m$, and a generalized diagonal matrix in the continuous
indices like $k$ \footnote{{E. g.: We can deal with this generalized matrix
rigging the space ${\cal S}$ and using the Gel'fand-Maurin theorem \cite
{Gorini}, this procedure allows us to define a generalized state eigenbasis
for system with continuous spectrum. It has been used to diagonalize
hamiltonians with continuous spectra in \cite{Bohm}, \cite{CGG}, \cite
{Laura1}, etc.}}{.} Under the diagonalization process eq. (\ref{4.1}) is
written as: 
\begin{equation}
(\rho _{*}|=\int U_{[{\bf k}]}^{\dagger [{\bf l}]}\rho _{\omega [{\bf l][l}%
^{\prime }]}U_{[{\bf k}^{\prime }]}^{[{\bf l}^{\prime }]}U_{[{\bf k}^{\prime
}]}^{\dagger [{\bf l}^{\prime \prime }]}(\omega ,[{\bf l}^{\prime \prime }%
{\bf ],[l}^{\prime \prime \prime }]|U_{[{\bf k}]}^{[{\bf l}^{\prime \prime
\prime }]}d\omega d[{\bf k]d[k}^{\prime }]d[{\bf l]d[l}^{\prime }]d[{\bf l}%
^{\prime \prime }{\bf ]d[l}^{\prime \prime \prime }]  \label{4.4'}
\end{equation}
where $U_{[{\bf k}]}^{\dagger [{\bf l}]}$ is the unitary matrix used to
perform the diagonalization and: 
\begin{equation}
\rho _{\omega [{\bf l][l}^{\prime }]}=\rho _{\omega [{\bf l]}}\delta _{[{\bf %
l][l}^{\prime }]}  \label{4.4''}
\end{equation}
where: 
\begin{equation}
\rho _{\omega [{\bf l][l}]}=\rho _{\omega [{\bf l]}}=\int U_{[{\bf l}]}^{[%
{\bf k}]}\rho _{\omega [{\bf k][k}^{\prime }]}U_{[{\bf l}]}^{\dagger [{\bf k}%
^{\prime }]}d[{\bf k}]d[{\bf k}^{\prime }]  \label{4.4'''}
\end{equation}
so we can define: 
\begin{equation}
(\omega ,[{\bf l]}|=(\omega ,[{\bf l],[l}]|=\int U_{[{\bf l}]}^{[{\bf k}%
]}(\omega ,[{\bf k],[k}^{\prime }]|U_{[{\bf l}]}^{\dagger [{\bf k}^{\prime
}]\dagger }d[{\bf k}]d[{\bf k}^{\prime }]  \label{4.4''''}
\end{equation}
We can repeat the procedure with vectors $(\omega ,\omega ^{\prime },[{\bf %
k],[k}^{\prime }]|$ and obtain vector $(\omega ,\omega ^{\prime },[{\bf l]|.}
$ In this way we obtain a diagonalized cobasis \{$(\omega ,[{\bf l]}%
|,(\omega ,\omega ^{\prime },[{\bf l]}\}.$ So we can now write the
equilibrium state as: 
\begin{equation}
\rho _{*}=\int \rho _{\omega [{\bf l]}}(\omega ,[{\bf l}]|d\omega d[{\bf l}]
\label{4.5}
\end{equation}
Since vectors $(\omega ,[{\bf l}]|$ can be considered as diagonals in all
the variables we have obtained decoherence in all the dynamical variables.
This fact will become more clear once we study the observables related with
this vector and introduce the notion of {\it pointer basis.}

So, let us now consider the observable basis \{$|\omega ,[{\bf l]}),|\omega
,\omega ^{\prime },[{\bf l])}\}$ dual to the state cobasis $\{(\omega ,[{\bf %
l]}|,(\omega ,\omega ^{\prime },[{\bf l]|}\}.$ From eq. (\ref{3.4}) and as
the $\omega $ does not play any role in the diagonalization procedure we
obtain: 
\begin{equation}
|\omega ,[{\bf l]})=|\omega ,[{\bf l}]\rangle \langle \omega ,[{\bf l}%
]|,\qquad |\omega ,\omega ^{\prime },[{\bf l])=}|\omega ,[{\bf l}]\rangle
\langle \omega ^{\prime },[{\bf l}]|  \label{4.5'}
\end{equation}
So in the basis \{$|\omega ,[{\bf l]}),|\omega ,\omega ^{\prime },[{\bf l])}%
\}$ the hamiltonian reads: 
\begin{equation}
h=\int \omega |\omega ,[{\bf l}])d\omega d[{\bf l}]=\int \omega |\omega ,[%
{\bf l}]\rangle \langle \omega ,[{\bf l}]|d\omega d[{\bf l}]  \label{4.6}
\end{equation}
Now, we can also define the operators: 
\begin{equation}
{\bf L}=\int {\bf l}|\omega ,[{\bf l}])d\omega d[{\bf l}]=\int {\bf l}%
|\omega ,[{\bf l}]\rangle \langle \omega ,[{\bf l}]|d\omega d[{\bf l}]
\label{4.7}
\end{equation}
that can also be written: 
\begin{equation}
L_i=\int l_i|\omega ,[{\bf l}])d\omega d[{\bf l}]=\int l_i|\omega ,[{\bf l}%
]\rangle \langle \omega ,[{\bf l}]|d\omega d[{\bf l}]  \label{4.8}
\end{equation}
where $i$ is an index such that it covers all the dimension of the ${\bf l}$
. Now we can consider the set $(h,L_i),$ which is a CSCO, since all the
members of the set commute, because they share a common basis and find the
corresponding eigenbasis of the set, precise $|\omega ,[{\bf l}]\rangle $
since \footnote{%
In some occasions we will call $h=L_0$ and $\omega =l_0.$}: 
\begin{equation}
h|\omega ,[{\bf l}]\rangle =\omega |\omega ,[{\bf l}]\rangle  \label{4.9}
\end{equation}
\begin{equation}
L_i|\omega ,[{\bf l}]\rangle =l_i|\omega ,[{\bf l}]\rangle  \label{4.10}
\end{equation}
Of course the $L_i$ are constant of the motion because they commute with $h.$

From all these equations we can say that:

i.- $(h,L_i)$ is the pointer CSCO.

ii.- \{$|\omega ,[{\bf l]}),|\omega ,\omega ^{\prime },[{\bf l])}\}$ is the
pointer observable basis.

iii.-$\{(\omega ,[{\bf l]}|,(\omega ,\omega ^{\prime },[{\bf l]|}\}$ is the
pointer states cobasis.

In fact, from eq. (\ref{4.5}) we see that the final equilibrium state has
only diagonal terms in this state (those corresponding to vectors $(\omega ,[%
{\bf l]}|)$ , it has not off-diagonal terms (those corresponding to vectors $%
(\omega ,\omega ^{\prime },[{\bf l]|,}(\omega ,[{\bf k],[k}^{\prime }]|,$ or 
$(\omega ,\omega ^{\prime },[{\bf k],[k}^{\prime }]|),$ and therefore we
have decoherence in all the dynamical variables.

\section{Correlations.}

As it was explained in paper \cite{LauraF} correlations are computed in the
limit of small $\hbar .$ Under this assumption it is demonstrated that for
each observable (e. g.: momenta or energy) we can find a canonically
conjugated dynamical variable (e. g.: configuration variables or the hand of
a clock, namely time), if we neglect $O(\hbar )$ terms. So we will use these
approximated canonically conjugated variables in this section, since, we
repeat, we are only interested in observational conditions where $\hbar $
can be considered very small.

Accordingly with this idea the canonically conjugated variable of $h$ would
be essentially $\eta ,$ but since $\rho _{*}$ is a $\eta $-constant, the
time variable is completely unimportant in this section (we will discuss
this matter further on section 6.3). Let us call $a_i$ the canonically
conjugated variable (precisely ''conjugated variable up to $O(\hbar )$
terms'') of the observable $L_i.$ Then, $(a_i)$ will be our configuration
variables and $(L_i$) our momentum variables. (We will call $x,p$ to the old
variables of eq. (\ref{2.1}) and ${\bf \xi }$ or ${\bf a}$ to the new
configuration variable and ${\bf \pi }$ or ${\bf l}$ to the new momentum
variables). Using these new variables we will compute the Wigner function 
\cite{Wigner} corresponding to the operator $\rho _{*.}$ We can also use the
usual transcription rules: 
\begin{equation}
h\rightarrow i\frac \partial {\partial \eta },\qquad L_i\rightarrow -i\frac 
\partial {\partial a_i}  \label{5.1}
\end{equation}
since the difference with respects to other transcription rules in other
coordinates is just a $O(\hbar ).$ Then: 
\begin{equation}
\langle \eta ,[\Delta {\bf a+a}_0]{\bf |}\omega [{\bf l]}\rangle
=e^{i(-\omega \eta +{\bf l\bullet \Delta a})}\langle 0,[{\bf a}_0]{\bf |}%
\omega [{\bf l]}\rangle  \label{5.2}
\end{equation}
But we will only consider the state of affairs for $\eta =0.$ We will call: 
\begin{equation}
\lbrack {\bf a}]=[\Delta {\bf a+a}_0],\qquad \{l\}=(\omega ,[{\bf l}%
]),\qquad \{\pi \}=(\omega ,[{\bf \pi ])}  \label{5.3}
\end{equation}
where, in the second and third equations we have restored the notation of
eq. (\ref{2.15}) With this notation and for $\eta =0$, eq. (\ref{5.2})
reads: 
\begin{equation}
\langle [{\bf a}]|\{l\}\rangle =e^{i\Delta {\bf a\bullet l}}\langle [{\bf a}%
]|\{l\}\rangle  \label{5.4}
\end{equation}
The Wigner function corresponding to matrix $\rho _{*}$ reads: 
\begin{equation}
F_{W*}(x,p)=F_{W*}({\bf \xi ,\pi })\sim \int_{-\infty }^\infty \langle {\bf %
\xi -\eta }|\rho _{*}|{\bf \xi +\eta }\rangle e^{2i{\bf \pi \bullet \eta }}d[%
{\bf \eta ]}  \label{5.5}
\end{equation}
Then from eqs. (\ref{4.5}) and (\ref{5.4}) (and eq. (\ref{6.7}), written for
the spatial coordinates for the continuous indices, see details in \cite
{LauraF}) we have: 
\[
F_{W*}(x,p)\sim \int ...\int_{-\infty }^\infty d[{\bf \eta ]}\int d\{l\}\rho
_{\{l\}}\langle {\bf \xi -\eta }|(\{l\}||{\bf \xi +\eta }\rangle e^{2i{\bf %
\pi \bullet \eta }}= 
\]
\[
\int ...\int_{-\infty }^\infty d[{\bf \eta ]}\int d\{l\}\rho _{\{l\}}e^{i(%
{\bf \xi -\eta })l}(\{l\}|[{\bf a}_0])e^{-i({\bf \xi +\eta })l}e^{2i{\bf \pi
\bullet \eta }}\sim 
\]
\begin{equation}
\int d\{l\}\rho _{\{l\}}|\langle \{l\}|[{\bf a}_0]\rangle |^2\delta ([{\bf %
\pi ]-[l])}  \label{5.6}
\end{equation}
where the probability $(\{l\}|[{\bf a}_0])$ has been called with the more
familiar (but not rigorous) symbol $|\langle \{l\}|[{\bf a}_0]\rangle |^2$
(that turns out to be rigorous only for the discrete $l)$ \footnote{%
We see that ${\bf \xi }$ disappears from the equation, so $F_{W*}({\bf \xi
,\pi })$ is neither a function of the position ${\bf \xi }$ nor of the
conventional origin ${\bf a}_0$. This is a consequence of the spatial
homogeneity of the model we are studying. Moreover, it can also be seen that
if we use eq. (\ref{5.2}) with $\eta \neq 0$ the function $F_{W*}({\bf \xi
,\pi })$ is a constant of $\eta $ (as it should be). So, essentially, in all
this section we are dealing with functions that are constants in time.}$. $
The $\delta ([{\bf \pi ]-[l])}$ does not contain the energy. But in the
footnote of section 6.3 we will prove that a $\delta -$term in the energy
can also be added, so finally: 
\begin{equation}
F_{W*}(x,p)\sim \int d\{l\}\rho _{\{l\}}|\langle \{l\}|[{\bf a}_0]\rangle
|^2\delta (\{\pi \}-\{l\})=\int d\{l\}\rho _{\{l\}}|\langle \{l\}|[{\bf a}%
_0]\rangle |^2\prod_{i=0}\delta (\pi _i-l_i)  \label{5.7}
\end{equation}
The last equation can be interpreted as follows:

i.- $\delta (\{p\}-\{l\})$ is a classical density function, strongly peaked
at certain values of the constants of motion $\{l\},$ corresponding to a set
of trajectories, where the momenta are equal to the eigenvalues of eqs. (\ref
{4.9}) and (\ref{4.10}), namely $\pi _i=l_i$ $(i=0,1,2,...)$. This fact
already shows the presence of correlations in our model. In fact: We can
consider each set of trajectories labelled by $\{l\}$ (i.e. a ''history''
obtained using some apparatuses that measure only the momenta) and prove
that in these trajectories the usual coordinate $x$ and the usual momentum $%
p $ are correlated as it is allowed by the uncertainty principle (see \cite
{LauraF}). For the conjugated variables ${\bf l}$ and ${\bf a,l}$ is
completely defined and ${\bf a}$ is completely undefined, satisfying also
the uncertainty principle.

ii.- $\rho _{\{l\}}$ is the probability to be in one of these sets of
trajectories labelled by $\{l\}$. Precisely: if some initial density matrix
is given, from eq. (\ref{4.5}) it is evident that its diagonal terms $\rho
_{\{l\}}$ are the probabilities to be in the states $(\omega ,[{\bf l]|}$
and therefore the probability to find, in the corresponding classical
equilibrium density function $F_{W*}(x,p)$, the density function $\delta
(\{p\}-\{l\}),$ namely the set of trajectories labelled by $\{l\}=(\omega ,[%
{\bf l])}.$

iii.- The factor $|\langle [{\bf a}_0]|\{l\}\rangle |^2$ corresponds to the
probability that one of the trajectories $\{l\}$ would pass by ${\bf a}_0$
at time $\eta =0$ and it can easily be computed from the model \footnote{%
From the spatial homogeneity of the problem and the usual normalization we
have $(\{l\}|[{\bf a}_0])=|\langle [{\bf a}_0]|\{l\}\rangle |^2\sim \omega
^{-n}$, being $n$ the particle number.}.

iv.- Therefore $\rho _{\{l\}}|\langle [{\bf a}_0]|\{l\}\rangle |^2=p_{\{l\}[%
{\bf a}_0]}$ is the probability that, given an initial density matrix, a
trajectory, with constant of the motion $\{l\}$ would pass by the point $%
{\bf a}_0$ at time $\eta =0$, and then it would follow the classical
trajectory: 
\begin{equation}
{\bf a=l}\eta {\bf +a}_0  \label{5.8}
\end{equation}
But, really $p_{\{l\}[{\bf a}_0]}$ is not a function of ${\bf a}_0$, it is
simply a constant in ${\bf a}_0$ (as explained in a previous footnote) since
this is only an arbitrary point and our model is spatially homogenous, we
can write: 
\begin{equation}
p_{\{l\}[{\bf a}_0]}=\int p_{\{l\}[{\bf a}_0]}\prod_{i=1}\delta (\xi
_i-a_{0i})d[{\bf a}_0]  \label{5.8'}
\end{equation}
in this way we have changed the role of ${\bf a}_0$, it was a fixed (but
arbitrary) point and it is now a variable that moves all over the space.
Then eq. (\ref{5.7}) reads: 
\begin{equation}
F_{W*}(x,p)\sim \int p_{\{l\}[{\bf a}_0{\bf ]}}\prod_{i=0}\delta (\pi
_i-l_i)\prod_{j=1}\delta (\xi _j-a_{0j})d[{\bf a}_0]d\{l\}  \label{5.8'''}
\end{equation}
So if we call : 
\begin{equation}
F_{W\{l\}[{\bf a}_0{\bf ]}}(x,p)=\prod_{i=0}\delta (\pi
_i-l_i)\prod_{j=1}\delta (\xi _j-a_{0j})  \label{5.8''''}
\end{equation}
we have: 
\begin{equation}
F_{W*}(x,p)\sim \int p_{\{l\}[{\bf a}_0{\bf ]}}F_{W\{l\}[{\bf a}_0{\bf ]}%
}(x,p)d[{\bf a}_0{\bf ]}d\{l\}  \label{5.8'''''}
\end{equation}
From eq. (\ref{5.8''''}) we see that $F_{W\{l\}[{\bf a}_0{\bf ]}}(x,p)\neq 0$
only in a narrow strip around the classical trajectory (\ref{5.8}) defined
by the momenta $\{l\}$ and passing through the point [${\bf a}_0{\bf ]}$
(really the density function is as peaked as it is allowed by the
uncertainty principle, so its width is essentially a $O(\hbar ),),$ since
the $\delta -$functions of all the equation are really Dirac's deltas when $%
\hbar \rightarrow 0)$ These trajectories explicitly show the presence of
correlations in our model \footnote{%
Of course, our ''trajectories'' are not only one trajectory for a one
particle state, but they are $n$ trajectories (each one corresponding to a
momenta $(l_1,l_2,...l_n)=\{l\}$ and passing by a point (${\bf a}_1,{\bf a}%
_2,...,{\bf a}_n)=[{\bf a])}$ for the n particle states. As $%
p_{\{l\}\{a\}}\sim \omega ^{-n}$ the probability decreases with the particle
number and the energy.}. So we have proved eq. (\ref{5.8'''''}) which, in
fact, it is eq. (\ref{1.1}) as announced \footnote{%
In this section we have faced the following problem:
\par
$F_{W*}({\bf \xi ,\pi })$ is a ${\bf \xi }$ constant that we want to
decompose in functions $F_{W\{l\}[{\bf a}_0{\bf ]}}(x,p)$ which are
different from zero only around the trajectory (\ref{5.8}) and therefore are
variables in ${\bf \xi .}$%
\par
Then,essentially we use the fact that if $f(x,y)=g(y)$ is a constant
function in $x$ we can decompose it as: 
\[
g(y)=\int g(y)\delta (x-x_0)dx_0 
\]
namely the densities $\delta (x-x_0)$ are peaked in the trajectories $%
x=x_0=const.,y=var.$ and, therefore, are functions of $x.$ This trajectories
play the role of those of eq. (\ref{5.8'}).
\par
As all the physics, including the correlations, is already contained in eq. (%
\ref{5.7}) (as explained in point i) the reader may just consider the final
part of this section, from eq. (\ref{5.8'}) to eq. (\ref{5.8'''''}) a
didactical trick.}.

Then we have obtained the classical limit. When $\eta \rightarrow \infty $
the quantum density $\rho $ becomes a diagonal density matrix $\rho _{*}.$
The corresponding classical distribution $F_{W*}(x,p)$ can be expanded as a
sum of classical trajectories density functions $F_{W\{l\}[{\bf a}_0{\bf ]}%
}(x,p),$ each one weighted by its corresponding probability $p_{\{l\}[{\bf a}%
_0{\bf ]}}.$ So, as the limit of our quantum model we have obtained a
statistical classical mechanical model, and the classical realm appears.

\section{Discussion and comments.}

\subsection{Characteristic times.}

The decaying term of eq. (\ref{3.8}) (i. e. the second term of the r. h. s.)
can be analytically continued using the techniques explained in papers \cite
{LauraA}, \cite{Cast}, and \cite{Laura1}. In these papers it is shown that
each pole $z_i=\omega _i-i\gamma _i,$ of the S-matrix of the problem
considered, originates a damping factor $e^{-\gamma _i\eta }.$ Then if $%
\gamma =\min (\gamma _i)$ the characteristic decoherence time is $\gamma
^{-1}.$ This computation is done in the specific models of papers \cite{Cast}%
. If $\gamma \ll 1,$ even if the Riemann-Lebesgue theorem is always valid,
there is no practical decoherence since $\gamma ^{-1}\gg 1.$

\subsection{Sets of trajectories decoherence.}

It is usual to say that in the classical regime there is decoherence of the
set trajectories labelled by the constant of the motion $\omega ,$ $[{\bf l}%
] $. This result can easily be obtained with our method in the following way.

i.- Let us consider two different states $|\omega [{\bf l}]\rangle $ and $%
|\omega ^{\prime }[{\bf l}^{\prime }]\rangle $ that will define classes of
trajectories with different constants of the motion $(\omega ,[{\bf l}])\neq
(\omega ^{\prime },[{\bf l}^{\prime }]).$ We must compute: 
\begin{equation}
\langle \omega [{\bf l}]|\rho _{*}|\omega ^{\prime }[{\bf l}^{\prime
}]\rangle =(\rho _{*}||\omega \omega ^{\prime }[{\bf l}][{\bf l}^{\prime
}])=\left[ \int \rho _{\omega ^{\prime \prime }[{\bf l}^{\prime \prime
}]}(\omega ^{\prime \prime }[{\bf l}^{\prime \prime }]|d\omega ^{\prime
\prime }d[{\bf l}^{\prime \prime }]\right] |\omega \omega ^{\prime }[{\bf l}%
][{\bf l}^{\prime }])=0  \label{6.1}
\end{equation}
due to the orthogonality of the basis $\{(\omega ,[{\bf l]}|,(\omega ,\omega
^{\prime },[{\bf l]|}\}$ .

ii.- But if we compute: 
\[
\langle \omega [{\bf l}]|\rho _{*}|\omega [{\bf l}]\rangle =(\rho
_{*}||\omega [{\bf l}])=\left[ \int \rho _{\omega ^{\prime \prime }[{\bf l}%
^{\prime \prime }]}(\omega ^{\prime \prime }[{\bf l}^{\prime \prime
}]|d\omega ^{\prime \prime }d[{\bf l}^{\prime \prime }]\right] |\omega
\omega [{\bf l}])= 
\]
\begin{equation}
\int \rho _{\omega ^{\prime \prime }[{\bf l}^{\prime \prime }]}\delta
(\omega -\omega ^{\prime \prime })\delta ([{\bf l}]-[{\bf l}^{\prime \prime
}])d\omega ^{\prime \prime }d[{\bf l}^{\prime \prime }]=\rho _{\omega [{\bf l%
}]}\neq 0  \label{6.2}
\end{equation}
The last two equations complete the demonstration. We will discuss the
problem of the decoherence of two trajectories, with the same $\{l\}$ but
different $[{\bf a}_0]$ in subsection 6.4.

\subsection{A discussion on time decoherence.}

It is well known that one of the main problems of quantum gravity is the
problem of the time definition (see \cite{Time}). A not well studied feature
of this problem is that, there must be a decoherence process related with
time, since time is as a classical variable. In this subsection, using the
functional technique, we will give a model that shows that this is the
case.(but we must emphasize that this subject is not completely developed).

We must compute $\langle \eta |\rho _{*}|\eta ^{\prime }\rangle $ where $%
|\eta \rangle $ and $|\eta ^{\prime }\rangle $ are two states of the system
for different times that evolve as \footnote{%
Cf. eq. (\ref{5.2}) and remember that therefore in this subsection we are
dealing with equations only valid when $\hbar \rightarrow 0.$}: 
\begin{equation}
|\eta \rangle =e^{-ih\eta }|0\rangle  \label{6.3}
\end{equation}
$|\eta \rangle \langle \eta ^{\prime }|$ can be considered as an observable,
then: 
\begin{equation}
\langle \eta ^{\prime }|\rho _{*}|\eta \rangle =(\rho _{*}||\eta \rangle
\langle \eta ^{\prime }|)  \label{6.4}
\end{equation}
But: 
\begin{equation}
(\omega ||\eta \rangle \langle \eta ^{\prime }|)=(\omega |e^{-ih\eta
}|0\rangle \langle 0|e^{ih\eta ^{\prime }})=[e^{ih\eta ^{\prime }}(\omega
|e^{-ih\eta }]||0\rangle \langle 0|)  \label{6.5}
\end{equation}
Now, for any observable $O$ we have: 
\[
\lbrack e^{ih\eta ^{\prime }}(\omega |e^{-ih\eta }]||O)=[e^{ih\eta ^{\prime
}}(\omega |e^{-ih\eta }]|[\int O_{\omega ^{\prime }}|\omega ^{\prime
})d\omega ^{\prime }+\int \int O_{\omega ^{\prime }\omega ^{\prime \prime
}}|\omega ^{\prime },\omega ^{\prime \prime })d\omega ^{\prime }d\omega
^{\prime \prime })= 
\]
\[
\lbrack e^{ih\eta ^{\prime }}(\omega |e^{-ih\eta }]|[\int O_{\omega ^{\prime
}}|\omega ^{\prime })d\omega ^{\prime }+...=(\omega |[\int O_{\omega
^{\prime }}e^{-i\omega ^{\prime }\eta }|\omega ^{\prime })e^{i\omega
^{\prime }\eta ^{\prime }}d\omega ^{\prime }]= 
\]
\begin{equation}
e^{-i\omega (\eta ^{\prime }-\eta )}(\omega |O)  \label{6.6}
\end{equation}
Thus \footnote{%
Considering this equation and repeating the procedure done, from eq. (\ref
{5.5}) to eq. (\ref{5.7}), we can see that there is an extra $\delta -$%
factor $\delta (\pi _0-l_0)$ related with the energy. Therefore the
trajectories described in section 5 conserve, not only the momenta $l$ , but
also the energy $h.$}: 
\begin{equation}
(\omega ||\eta \rangle \langle \eta ^{\prime }|)=e^{-i\omega (\eta ^{\prime
}-\eta )}(\omega ||0\rangle \langle 0|)  \label{6.7}
\end{equation}
So now we can compute the following two cases:

i.- 
\begin{equation}
\langle \eta ^{\prime }|\rho _{*}|\eta \rangle =(\rho _{*}||\eta \rangle
\langle \eta ^{\prime }|)=[\int \rho _\omega (\omega |d\omega ]||\eta
\rangle \langle \eta ^{\prime }|)=\int \rho _\omega e^{-i\omega (\eta
^{\prime }-\eta )}(\omega ||0\rangle \langle 0|)d\omega \rightarrow 0
\label{6.8}
\end{equation}
when%
\mbox{$\vert$}
$\eta ^{\prime }-\eta |\rightarrow \infty $ , due to the Riemann-Lebesgue
theorem$.$

ii.- Analogously:

\begin{equation}
\langle \eta |\rho _{*}|\eta \rangle =\int \rho _\omega (\omega ||0\rangle
\langle 0|)d\omega \neq 0  \label{6.9}
\end{equation}
So we have time decoherence for two times $\eta $ and $\eta ^{\prime }$ if
they are far enough.

This result is important for the problem of time definition, since in order
to have a reasonable classical time this variable must first decohere. The
result above shows that this is the case for $\eta $ and $\eta ^{\prime }$
far enough, but also that, for closer times (namely such that their
difference is smaller than Planck's time) there is no decoherence and time
cannot be considered as a classical variable. Classical time is a familiar
concept but the real nature of the non-decohered quantum time is opened to
discussion.

\subsection{Decoherence in the space variables.}

Now that we know that there is time decoherence we can repeat the reasoning
for the rest of the variables ${\bf \xi }$ at time $\eta =0$ and changing
eq. (\ref{6.3}) by: 
\begin{equation}
|{\bf \xi \rangle =}e^{i{\bf \xi \bullet l}}|{\bf 0\rangle }  \label{6.10}
\end{equation}
and we will reach to the conclusions:

i.- 
\begin{equation}
\langle {\bf \xi }|\rho _{*}|{\bf \xi }^{\prime }\rangle \rightarrow 0
\label{6.11}
\end{equation}
when 
\mbox{$\vert$}
${\bf \xi -\xi }^{\prime }|\rightarrow \infty .$

ii,- 
\begin{equation}
\langle {\bf \xi }|\rho _{*}|{\bf \xi }\rangle \neq 0  \label{6.12}
\end{equation}
therefore there is also decoherence between two trajectories with the same $%
\{l\}$ but different $[{\bf a}_0]$.

These facts complete the scenario about decoherence and correlations.

\section{Conclusion.}

After the WKB expansion and the decoherence and correlations process our
quantum model has:

i.-A defined classical time $\eta $ and a defined classical geometry related
by eq. (\ref{2.10}).

ii.- Decoherence has appeared in a well defined pointer basis.

iii.- The quantum field has originated a classical final distribution
function (eq. (\ref{5.8'''''})) that is a weighted average of some set
densities, each one related to a classical trajectory. The weight
coefficients are the probabilities of each trajectory.

We can foresee that if instead of a spinless field we would coupled the
geometry with a spin 2 metric fluctuation field the result would be more or
less the same. Then the corresponding quantum fluctuations would become
classical fluctuations that would correspond to matter inhomogeneities
(galaxies, clusters of galaxies, etc.) that will move along the trajectories
described above. But this subject will be treated elsewhere with great
detail.

\section{Acknowledgments.}

This work was partially supported by grants CI1$^{*}$-CT94-0004 and PSS$%
^{*}-0992$ of the European Community, PID 3183/93 of CONICET, EX053 of the
Buenos Aires University, and also grants from Fundaci\'{o}n Antorchas and
OLAM Foundation.

\end{document}